\documentclass[prl,twocolumn,showkeys,preprintnumbers,amsmath,amssymb]{revtex4}

\usepackage{natbib}
\usepackage{graphicx}
\usepackage{dcolumn}
\usepackage{bm}

\newcommand{\ket}[2][]{{|#2\rangle_{#1}}}
\newcommand{\bra}[2][]{{}_{#1}\langle #2|}

\newcommand{\Tr}{\textrm{Tr}}
\newcommand{\wt}{\widetilde}

\def\duzomniejsze{<\kern-.7mm<}
\def\duzowieksze{>\kern-.7mm>}

\def\textbf#1{{\bf #1}}
\def\beq{\begin{equation}}
\def\eeq{\end{equation}}
\def\be{\begin{equation}}
\def\ee{\end{equation}}
\def\ben{\begin{eqnarray}}
\def\een{\end{eqnarray}}
\def\beqa{\begin{eqnarray}}
\def\eeqa{\end{eqnarray}}
\def\eea{\end{array}}
\def\bea{\begin{array}}
\newcommand{\bei}{\begin{itemize}}
\newcommand{\eei}{\end{itemize}}
\newcommand{\bee}{\begin{enumerate}}
\newcommand{\eee}{\end{enumerate}}

\newcommand{\p}{\mathrm{P}} 

\usepackage{xcolor}
\newcounter{mnotecount}[section]
\renewcommand{\themnotecount}{\thesection.\arabic{mnotecount}}
\newcommand{\mnote}[1]
{\protect{\stepcounter{mnotecount}}$^{\mbox{\footnotesize
$
\bullet$\themnotecount}}$ \marginpar{
\raggedright\tiny\em
$\!\!\!\!\!\!\,\bullet$\themnotecount: #1} }

\usepackage{soul}

\usepackage{hyperref}

\begin{document}


\title{Quantum black holes as solvents}

\author{Erik Aurell}
\email{eaurell@kth.se}
\affiliation{KTH -- Royal Institute of Technology, AlbaNova University Center, SE-106 91 Stockholm, Sweden}%
\affiliation{
Faculty of Physics, Astronomy and Applied Computer Science, Jagiellonian University, 30-348 Krak\'ow, Poland
}

\author{Micha{\l} Eckstein}
\email{michal@eckstein.pl}
\affiliation{Institute of Theoretical Physics,
Jagiellonian University, ul. {\L}ojasiewicza 11, 30--348 Krak\'ow, Poland}
\affiliation{Institute of Theoretical Physics and Astrophysics, National Quantum Information Centre, Faculty of Mathematics, Physics and Informatics, University of Gda\'nsk, Wita Stwosza 57, 80-308 Gda\'nsk, Poland}

\author{Pawe{\l} Horodecki}
\email{pawhorod@pg.edu.pl}
\affiliation{International Centre for Theory of Quantum Technologies,
Uniwersytet Gda\'nski, 80-308 Gda\'nsk, Wita Stwosza 63, Poland}%
\affiliation{Faculty of Applied Physics and Mathematics, National Quantum Information Centre, 
Gda\'nsk University of Technology, Gabriela Narutowicza 11/12, 80-233 Gda\'nsk, Poland}

\date{24 April 2021}

\begin{abstract}
Almost all of the entropy in the universe is in the form of Bekenstein--Hawking (BH) entropy of super-massive black holes.
This entropy, if it satisfies Boltzmann's equation $S=\log{\cal N}$, hence represents almost 
all the accessible phase space of the Universe, somehow associated to objects which themselves
fill out a very small fraction of ordinary three-dimensional space.
Although time scales are very long, it is believed that black holes will eventually evaporate by emitting Hawking radiation, which is thermal when counted mode by mode.
A pure quantum state collapsing to a black hole will hence eventually re-emerge as a state with strictly positive entropy, which constitutes the famous black hole information paradox. Expanding on a remark by Hawking we posit that BH entropy is a thermodynamic entropy, which must be distinguished from information-theoretic entropy. 
The paradox can then be explained by information return in Hawking radiation.
The novel perspective advanced here is that if
BH entropy counts the number of accessible physical states in a quantum black hole, then the paradox can be seen as an instance of the fundamental problem of statistical mechanics. We suggest a specific analogy to the increase of the entropy in a solvation process.
We further show that the huge phase volume (${\cal N}$), which must be made available to the universe in a gravitational collapse, cannot originate from the entanglement between ordinary matter and/or radiation inside and outside the black hole. We argue that, instead, the quantum degrees of freedom of the gravitational field must get activated near the singularity, resulting in a final state of the `entangled entanglement' form involving both matter and gravity.
\end{abstract}

\maketitle

\section{Black hole information paradox}

The black hole information paradox can be formulated as the following \textit{Gedankenexperiment}.
Assume that quantum mechanics is universally valid, that gravitation
is switched off, and that a sufficiently large amount of matter can be found
in a pure quantum state. If gravitation is switched on this matter
will collapse into a black hole, which by unitarity will still be in a pure
quantum state.
At a later time this state will evolve to the remainder of the black hole and the Hawking radiation escaping to infinity. Taken separately, neither
the remainder black hole nor the escaping radiation are pure; the black
hole state is supposed to tend to a maximally mixed state on all black hole states with the same mass, and
the radiation is assumed to be thermal. 
Then, when the black hole evaporation
reaches its end, there is no black hole left, but only the radiation, which by assumption has strictly positive entropy.
Hence a pure state has evolved into a mixed state, which would break the unitary evolution in quantum mechanics \cite{Hawking76,Hawking16}.
The topic has been reviewed multiple times, recently in \cite{Verlinde-Verlinde2013,Harlow,Marolf} and \cite{Maldacena2020}.
The basic proposals for resolving the paradox
were summarized in \cite{Giddings98} or, more recently, \cite{UnruhWald17}
as fundamental information loss, remnants, and
information return in the Hawking radiation.
In this context ``firewalls''
and other scenarios involving new physics at the horizon fall into the
class of remnants since information is not destroyed and also does not leave
the vicinity of the black hole.

In this paper we propose a resolution in the direction
of information return, 
but in the framework of open quantum systems. 
Our focus here is not on how information 
actually returns in the Hawking radiation,
a topic to which we hope to return in the future, but 
what are its implications on the concept of entropy. 
A previous argument in the same vein was voiced by Stephen Hawking in \cite{Hawking14} but, to our best knowledge, it had no resonance in the literature. 
Part of our motivation in this paper is hence to bring attention to this
note by Hawking, and to place it in the
context of discussions of fundamental concepts of statistical mechanics.

The literature on quantum black holes is very large, and it is not possible to
refer to all relevant contributions in an regular article.
To guide the reader we here provide a partial summary, for 
more extensive discussions we refer to the recent reviews \textit{e.g.} 
\cite{Verlinde-Verlinde2013,Harlow,Marolf}.
First, as we pursue an approach of information return 
we will throughout assume
that quantum mechanics is universally valid.
This means that we have to assume that gravitation is also quantised.
In our proposal of quantum
entanglement in black hole formation
we assume, however, that without black holes the gravitational field is 
not entangled with other quantum systems.
From this standpoint the black hole information paradox 
is explained, at the kinematic level, by entanglement between the gravitational field
and other degrees of freedom during the collapse process.
As there is no 
established and generally accepted theory of quantum gravity
we adopt a theory-independent approach (cf., for instance, \cite{BellNL,Popescu14}). In other words, rather than providing an explicit physical mechanism, which would provide much more detailed understanding, we 
try to separate out the features which any theoretical model of this kind must have. We hence leave the interesting questions on the gravitational side, in particular those related to the localization of the Hawking radiation \cite{Hotta15,Wald19,Wilczek20}, for a future study.

The possibility of information return was first proposed by Page
\cite{Page1980}, and later developed by him in several publications, \textit{e.g.}~\cite{Page1993}.
The core problem is that if information is preserved, then the final state of the 
Hawking radiation, when the entire black hole has radiated away, has to be pure.
Since the states of all separate modes of Hawking radiation are thermal
this raises a compatibility issue between the total purity and the reduced mixed
states (cf. \cite{PaperII}). Furthermore, the difference must be encoded in quantum correlations
between the separate Hawking particles. But this is problematic, because these are, for the most part,
emitted one by one \cite{Gray16} and do not interact directly.
The related issues are discussed to this day: some 
notable contributions were made by Preskill \cite{Preskill1992}
(and later), Braunstein, Pirandola and \.{Z}yczkowski \cite{Braunstein2013};
Bradler and Adams \cite{BradlerAdami2014}; Lochan and Padmanaban \cite{Padmanabhan16}; Grudka and collaborators \cite{Grudka17};
and very recently by Marolf and Maxfield \cite{MarolfMaxfield2020}.
More contributions in these directions will be cited further on.

In this paper we will discuss the black hole information problem, under the assumption of a global unitary evolution, 
as an instance of what is often called the ``fundamental problem of statistical mechanics''. The latter is the question, dating back to Boltzmann, on how 
conservative classical dynamics can be reconciled with the increase of entropy and tendency to thermalization, in other words, 
with the Second Law. The quantum black hole problem has unusual features pertaining to the fact that the
phenomena involved are quantum and that the process itself involves unknown
constituents. The latter is so since a quantum black hole presumes quantum gravity in the strong-field regime near the singularity, and there is no
generally accepted theory for such phenomena. However, we find nonetheless that general
arguments of statistical mechanics do shed light on these issues, as first noted by Hawking \cite{Hawking14}.  

In ordinary room-temperature physics a low-entropy crystalline solid (solute) dissolves
into a high-entropy liquid (solvent), in the process called solvation. 
We propose that the massive increase in the entropy during the gravitational collapse is a physical process analogous to solvation, where the singularity dissolves the degrees of freedom of the gravitational field.

The paper is organized as follows.
In Section~\ref{sec:entropy-increase} we survey well-known results on the
Bekenstein--Hawking entropy, emphasizing the fact that it is by far the largest part of entropy in the Universe.
In Section~\ref{sec:entanglement} we present our proposal of considering the
Bekenstein--Hawking entropy as the entanglement entropy between the black hole
and the quantum gravitational field, and work out some simple consequences
of such a hypothesis \footnote{After completion of this work we found that mechanisms involving entanglement between matter and gravitational field have recently been proposed by Requardt \cite{Requardt} and Kay \cite{Kay}, but from different starting points, and with little overlap with our approach.}.
In Section~\ref{sec:foundational} we present our second argument about the
relation of the black hole information paradox to the foundational problem of statistical mechanics,
and in Section~\ref{sec:quantum-classical} we sum up and discuss the results.

\section{The entropy increase}
\label{sec:entropy-increase}

The Bekenstein--Hawking (BH) entropy of a black hole is
$S_{\mathrm{BH}}=k_B \frac{1}{4}\frac{A}{l_{\mathrm{P}}^2}$,
where $A$ is the surface area of the horizon and $l_{\mathrm{P}}=\sqrt{\frac{\hbar G}{c^3}}$ is the Planck length ($1.6\cdot 10^{-35}\hbox{m}$). The Boltzmann's constant $k_B = 1.38\cdot 10^{-23} \frac{J}{K}$ can be set to one by choosing a suitable temperature scale. In the following we will, however, keep it explicit.

Most astrophysical black holes are approximately stationary, have low rotation, as compared 
to the extreme allowed by the Kerr geometry, and are electrically neutral.
The space-time around such a black hole is thus well approximated by the spherically 
symmetric Schwarzschild geometry, where the   
radius of the trapping horizon is determined by its mass, $R=\frac{2GM}{c^2}$. The latter can also be written as $R=2 l_{\mathrm{P}}\frac{M}{m_{\p}}$
where $m_{\p}=\sqrt{\frac{c\hbar}{G}}$ is the Planck mass ($2\cdot 10^{-8}\hbox{kg}$).
BH entropy of a Schwarzschild black hole
is therefore also $S_{\mathrm{BH}}=k_B 4\pi\frac{M^2}{m_{\p}^2}$.
In ordinary thermodynamics entropy is an {\it extensive} quantity
\textit{i.e.} proportional to mass. The black hole entropy is, on the
other hand, {\it super-extensive}: it increases faster than linearly
with the mass. 
One simple consequence of this fact is the following: denote by $E_{\mathrm{BH}}=Mc^2$ the energy of a black hole. According to a standard thermodynamical formula the inverse temperature
is the partial derivative of entropy 
with respect to energy, $\frac{1}{T}= \frac{\partial S}{\partial E}$.
After inserting $S_{\mathrm{BH}}$ and $E_{\mathrm{BH}}$ from above we
obtain $\frac{\partial S_{\mathrm{BH}}}{\partial E_{\mathrm{BH}}}=\frac{4\pi k_B}{m_{\p}^2 c^2}\frac{\partial M^ 2}{\partial M}$. This gives a temperature, which is not a free parameter, but which
decreases with mass as  $T_{\mathrm{BH}}=\frac{m_{\p} c^2}{8\pi k_B} \left(\frac{M}{m_{\p}}\right)^{-1}$.
This temperature, which is the temperature of the Hawking radiation,
expressed in the units of Planck energy ($m_{\p} c^2 \approx 2\cdot 10^9 J$), is
weighted down by the mass of the black hole in Planck mass units.

It is also well established that BH entropy increases 
faster with mass than any
matter or light which
could have formed the black hole \cite{Bekenstein81}.
For example, 
for a photon gas we would have $S\sim M^{\frac{3}{2}}$ \cite{tHooft93,Harlow}, while $S_{BH} \sim M^2$.
Hence, it is clear that the BH entropy is a very special type of entropy,
probably unique in our Universe.
Furthermore, it not only increases faster with mass than anything else,
it is also in total much larger than anything else.
Since the BH entropy is super-extensive most of this type of entropy should be associated with the largest objects in the Universe. The largest such objects known are the
super-massive black holes found in the center of galaxies;
the fraction of the Universe's entropy \textit{not} inside a super massive black hole 
has been estimated to be as small as one part in $10^7$, cf. Table 1 in \cite{EganLineweaver10}.

In Boltzmann's theory of thermodynamics, entropy is a measure of the available
phase space $S=k_B \log{\cal N}$.
It is startling and perhaps paradoxical that super-massive black holes,
which occupy a very small volume of our Universe seem to contain
almost all the available phase space. If this is so, it is crucial to understand what is this phase space and how it
becomes accessible to the Universe via the creation of a black hole.

\section{Entanglement in black hole formation}
\label{sec:entanglement}
Consider an astrophysical black hole formed from some part of the universe which was
in a mixed state as far back as we can know.
Before the collapse it can hence be represented as a part of a larger pure state
\begin{equation}
  \label{eq:U=S+A}
\ket{\Psi^{(0)}}_U = \sum_i a_i \ket{i}_S\ket{i}_A,
\end{equation}
where $U$ stands for ``universe'', $S$ stands for ``system'' (or ``star'') and 
$A$ stands for the rest of the universe that here plays the role of an ``ancilla''. 
The von Neumann entropy of $S$ is the same as the entropy of $A$,
\begin{align}\label{SA}
S\Big[ \Tr_A \Big( \ket{\Psi^{(0)}}_U \bra{\Psi^{(0)}}_U \Big) \Big] & = - \sum_i |a_i|^2 \log |a_i|^2 \\
& = S\Big[ \Tr_S \Big( \ket{\Psi^{(0)}}_U \bra{\Psi^{(0)}}_U \Big) \Big], \notag
\end{align}
and it measures the entanglement between $S$ and $A$.
After the formation of the black hole one would usually expect to obtain the state
\begin{equation}
    \label{eq:U=BH+A}
\ket{\Psi^{(1)}}_U = \sum_i b_i \ket{i}_{BH}\ket{i}_{A'},
\end{equation}
where $A'$ stands for the rest of the universe outside the black hole,
and where the entropy of the black hole ($BH$) is much larger than the entropy of $S$. The gravitational collapse is essentially a local phenomenon, which starts when a (sufficiently massive) star losses its thermal equilibrium. Hence, the ``universe'' here means some neighborghood of the initial star, the size of which is limited by the finite speed of propagation of signals and physical interactions.

It is a fundamental result of quantum information theory
that entanglement cannot increase under local operations and classical communication (LOCC) \cite{Horodecki4}. 
This means that two experimentalists who operate on spacelike-separated parts of a
system, and who do not share a quantum resource and can only communicate classically to one another, cannot increase the entanglement between the two parts, whatever they do.
They can only decrease it. In fact, they can trivially eliminate the entanglement almost completely
by letting one or both parts decohere in contact with a bath.
In order to increase entanglement the two experimentalists and the system must act as one large 
quantum system. 
Similarly in natural systems, entanglement between two parts can only increase if these
parts interact, either directly, or through an intermediary.

For the entropy of the system to increase massively, the quantum matter forming the black hole 
would therefore have to massively interact
with the rest of the universe during the collapse process, and it cannot
do so with the surrounding ordinary matter and light through accretion disks, radiation, etc.
since all these would bring in only a tiny fraction of the required entropy.
In other words, the von Neumann entropy of a black hole, \textit{i.e.}  
\begin{align}\label{SBH}
S(\rho_{BH}) & = - \sum_i |b_{i}|^{2} \log |b_{i}|^{2}, \\
\text{ with } \; \rho_{BH} &  = \Tr_{A'} \Big( \ket{\Psi^{(1)}}_U \bra{\Psi^{(1)}}_U \Big) \notag
\end{align}
should not be very much larger than the von Neumann entropy 
of the initial system \eqref{SA}.
Since in fact BH entropy is
much greater than the entropy of the 
collection of the initial constituents, we must assume that the local unitary transformations involve
not only the constituents themselves but also some additional local parts 
carrying extra quantum degrees of freedom. 
Though theoretical models here greatly outweigh the facts,
it appears reasonable to assume that these unknown parts 
are related to the quantumness of the gravitational field, \textit{cf.}~\cite{Sorkin1997}. Such an assumption could be studied from various perspectives, for instance string theory \cite{Susskind94,Maldacena2020}, loop quantum gravity \cite{Perez15,Perez17,Perez19} or analogue gravity \cite{Entropy19}. Here we adopt a model-independent approach, basing solely on the general principles of quantum information (cf. \cite{NoHiding}). 

\begin{figure}
\centering
		\includegraphics[scale=1.15]{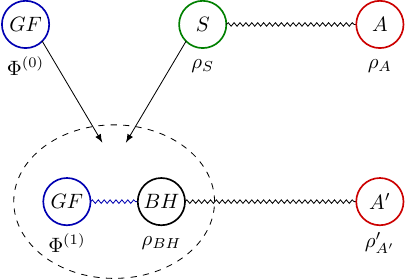}
		\caption{\label{fig}  The suggested picture of the process of black hole formation involving 
entanglement with the quantum degrees of freedom of the gravitational field,
initially assumed to be in a pure state $|\Phi^{(0)}\rangle_{GF}$ (see text).
Both the system $S$, i.e. the constituents that are to collapse, as well as the rest of the universe $A$,
have small initial von Neumann entropies $S(\rho_{S})=S(\rho_{A})= - \sum_{i} |a_{i}|^{2} \log |a_{i}|^{2}$.
Since the gravitational collapse is a local process it can be seen as a local unitary operation $U$ which leads to the interaction of
the system of constituents, the gravitational field and some limited part of the rest of the universe.
The final state representing entangled entanglement results in the black hole's entropy \eqref{SBH} 
much higher than the initial entropy of the constituents \eqref{SA}. 
This is precisely the effect of interaction with
the gravitational field which, initially, was not entangled with the other degrees of freedom. }
\end{figure}

Consequently, we should extend the initial quantum description \eqref{eq:U=S+A} to encompass the quantum degrees of freedom of the gravitational field. 
We assume that initially the gravitational field is not appreciably entangled with the
other degrees of freedom.
By linearity of quantum mechanics it then suffices to consider 
the case when the gravitational field is pure.
Hence, instead of \eqref{eq:U=S+A} the full initial quantum state before the collapse is
\begin{equation}
\ket{\wt{\Psi}^{(0)}}_{U+GF} = \ket{\Psi^{(0)}}_U \ket{\Phi^{(0)}}_{GF},
\end{equation}
where $GF$ stands for the gravitational field. 
Now, the local unitary evolution introduced above does not, by assumption, change the 
entropy of the outside part of the universe $A'$. On the other hand, it can 
 contribute to the massive entropy increase in a black hole by an interaction that 
leads to the final state:
\begin{equation}
\ket{\wt{\Psi}^{(1)}}_{U+GF} = \sum_i b_i \Big[ \sum_k c_{k}^{i}
\ket{k}_{BH} \ket{k}_{GF} \Big] \ket{i}_{A'}.
\end{equation}
States of this type are known in quantum information theory as
``entangled entanglement'' \cite{Zeilinger1996,EntEnt}.
Contrary to ordinary entanglement which involves one sum over
a Hilbert--Schmidt decomposition of a bipartite state, there is now
a double sum. In consequence the entanglement between any two subsystems is a property, which itself is entangled with the third subsystem. 
In the case at hand
the internal $BH+GF$ entanglement, conditioned at least upon
some of the states of the rest of the Universe $A'$, can then be much greater
than the one between the joint system $BH+GF$ and the rest of the
Universe $A'$ -- for an illustration, see Fig.~\ref{fig}. Indeed, with
\begin{align}\label{rhoBH}
\widetilde{\rho}_{BH} & = \Tr_{A'+GF} \Big( \ket{\wt{\Psi}^{(1)}}_{U+GF} \bra{\wt{\Psi}^{(1)}}_{U+GF} \Big) \\
& = - \sum_i |b_i|^2 \sum_k |c_k^i|^2 \ket{k}{}_{BH} \bra{k}_{BH} \notag
\end{align}
we have
\begin{align}\label{SBH_GF}
S(\widetilde{\rho}_{BH}) & = - \sum_{k,i} |b_i|^2 |c_k^i|^2 \log \Big( \sum_{j} |b_j|^2 |c_k^j|^2 \Big). 
\end{align}

The value of $S(\widetilde{\rho}_{BH})$ can easily become much larger than that of the initial configuration given by \eqref{SA}. Roughly, one can estimate (by assuming uniform distributions, $|a_i|^2 = 1/d_S$ and $|c_k^i|^2 = 1/d_{GF}$) that $S(\rho_{S}) \sim \log d_{S}$ and $S(\widetilde{\rho}_{BH}) \sim \log d_{GF}$, with $d_{S}$ and $d_{GF}$ denoting the numbers of active modes of the star and the gravitational field, respectively. Hence, the price to pay for avoiding the black hole information paradox is to assume that the gravitational collapse activates a huge number, $d_{GF} \gg d_S$, of quantum gravitational degrees of freedom, which remain idle if there is no black hole.

Finally, let us note that in the process of the black hole evaporation, the quantum correlations in the $BH+GF$ system can be transferred to the entanglement of the combined $BH+GF+HR$ system, where $HR$ stands for the Hawking radiation.
Eventually, when the black hole has radiated away completely, as it would do in a scenario with information return, there is only the Hawking radiation left, the modes of which are entangled.
The question of how this can actually happen is, however, outside of the scope of the present work. For some further kinematic
considerations, see \cite{PaperII}.

\section{Black hole information paradox as an instance of the foundational problem of statistical mechanics}
\label{sec:foundational}
If the BH entropy satisfies Boltzmann's formula ($S=\log{\cal N}$), then
the formation of a black hole makes
accessible to the universe a number
${\cal N} \sim e^{4\pi\frac{M^2}{m_{\p}^2}}$ of
states which were previously inaccessible.
In the language of quantum information
$S_{\mathrm{BH}}$ would be the log of the dimension of the subspace of Hilbert space needed to describe the interior
of a black hole.   
General arguments have been advanced against this so-called strong interpretation of BH entropy,~cf.~\cite{Hossenfelder2010}.

However, such arguments are somewhat troublesome in the light 
of modern statistical mechanics which strongly favours the objective interpretation
where entropy is associated with real degrees of freedom, and it is not only a measure of our ignorance.
Many papers on black hole thermodynamics and information paradox are written in the subjective tradition of thermodynamics founded on the max-entropy principle by Edwin T. Jaynes \cite{Jaynes1,Jaynes2}. The latter is invoked to draw conclusions about a complex system from a limited number of observables. Yet, Jaynes' principle was shown to lead to incorrect predictions when the studied phenomenon is not in equilibrium \cite{Vulpiani17}, which is certainly the case of the gravitational collapse. Furthermore, Jaynes' principle leads to wrong conclusions about the entanglement \cite{Horodecki99}, and must hence be considered incompatible with modern quantum information theory, where entanglement is fundamental \cite{Horodecki4}.  

In order to make a connection with thermodynamics, we first need to
distinguish between entropy in the
thermodynamic and the information-theoretic sense.
In the first case, 
following \cite{Hawking14}, BH entropy is assumed to be the log of the number of 
quantum states of the black hole with the same mass. 
In the second case, as discussed above around formulas
\eqref{eq:U=S+A} and \eqref{eq:U=BH+A}, BH entropy is the von Neumann entropy
measuring entanglement.
The perceived problem is that the first can be much larger than the second.
This can, however, be seen as
a reformulation of the foundational problem
of statistical mechanics
as posed by Thompson, Maxwell and Boltzmann \cite{Lebowitz1999}.
In other words, the problem is that the conservative classical dynamics preserves
entropy while many systems appear to behave as if tending to thermodynamic equilibrium
and then the entropy increases --- the Second Law applies.
A widely studied analogous problem is how high-dimensional quantum systems 
can behave as if they thermalize, the ``eigenstate thermalization hypothesis''
\cite{Deutch1991,Srednicki1994,Rigol2008,Reimann2013}.
For a further recent study in an explicitly gravitational context, see \cite{Oltean2016}.
An ensemble of classical deterministic systems will,
if the dynamics can be followed exactly, preserve Shannon entropy
(von Neumann entropy in the quantum case),
but will nevertheless after relaxation behave as if in thermal equilibrium, 
in an ensemble with the thermodynamic entropy.
As argued by Khinchin \cite{Khinchin,Vulpiani} the
systems under consideration in statistical mechanics
have very high dimensionality, and the kinds of observations that can be
made on them are simple. Both these properties hold for black holes:
the dimensionality is extremely high, and the
observations (or undisputed theoretical predictions) that can be made are quite limited. 

There is thus no paradox in the fact that the thermodynamic entropy of the
Hawking radiation, estimated by summing the contributions of all the modes, 
is much larger than the von Neumann entropy of the Hawking radiation. 
For a black hole that starts out in a pure state and eventually evaporates,
the time-evolution of its von Neumann entropy follows
the Page curve \cite{Page1993,Page2013} and finally drops to zero.

The total process whereby a collapsing star (in a pure quantum state)
eventually turns into photons and other Hawking particles streaming
away to infinity can be unitary (given an appropriate theory of quantum gravity)
and would then not increase the von Neumann entropy, but would be indistinguishable
(by the measurements that can practically be made)
from a process that actually does increase entropy, and which actually
is not unitary. 
Indeed, an observer independent from the system can effectively be coupled only to a few (collective) degrees
of freedom. In order to observe that the actual von Neumann entropy of the system
in question is small one would have to couple to {\it (almost) all}  degrees of freedom and then
unentangle them (i.e. effectively diagonalise the
system's density matrix) -- either via tomography and subsequent extensive (classical) post-processing, or physically
-- by accumulating the small von Neumann entropy
on a small part of the degrees of freedom, leaving
the others pure. The latter process is a generalised version, involving memory effects, of the standard concept of algorithmic cooling \cite{Schumacher96}. Hence, the observer would become an analogue of Maxwell's daemon who resets the temperature to much a lower value, corresponding to the von Neumann entropy~\cite{Vedral2000}. 

The novelty of our argument, which was previously stated \textit{en passant} by Hawking in
\cite{Hawking14}, is to point out that this
process also occurs in ordinary matter and has been at the forefront 
of research in statistical mechanics since the 19th century.
While quantum black holes hold many other mysteries, this is not one of them.

\section{Quantum vs classical gravity}
\label{sec:quantum-classical}

It is commonplace to assume that the resolution of the black hole information paradox requires a major departure from unitary quantum mechanics \cite{UnruhWald17} and/or the breakdown of the structure of relativistic spacetime \cite{Verlinde11,Grudka17}.
Our conclusion is that the information contained in the
infalling matter is not lost, but remains concealed in the quantum degrees of freedom
of the gravitational field which was initially not entangled with the matter.
At the same time those quantum degrees of freedom must be very thoroughly mixed
with those of the matter which fell into the black hole earlier.
Consequently, it is indeed possible for a quantum black hole 
to have a very large thermodynamic entropy, much larger than the entanglement entropy 
(von Neumann entropy) of all the initial matter.

The picture is analogous to the process of solvation: the black hole dissolves the solute --- the infalling matter along with the corresponding gravitational field, which have relatively low entropy. The resulting solution
will, on the other hand, have a relatively huge entropy. The very large increase in the entropy suggests that
solubility is very large
--- the solute is practically infinitely soluble in the solvent.

We should confront the present analysis with the fact that the gravitational field, say that of the Earth, acts on the particles
unitarily via an external potential. An analogous issue occurs in quantum optics: we know that whereas strong laser fields are fundamentally of quantum nature, they act on test quantum particles as effectively classical fields, via a time-dependent Hamiltonian. The fundamental difference is that the gravitational field, which affects the quantum
particles in a classical way is not strong at all. 
One could claim that big masses emit a large number of gravitons, corresponding to the quantum degrees of freedom of the field, which decohere on the way to the object. If there is a sufficiently
large number of them (eg. macroscopic objects) the internal nature
of the quantum gravitational field would lead to the emission of non-virtual
long-living quanta, which could travel for long distances.

A related issue is the expectation
\cite{BoseEtAl,MarlettoVedral} 
that small particles may entangle with each other via the gravitational
field (cf. \cite{Brukner2015,Belenchia2018,Belenchia2019,Zych2019}). In this case, the gravitons are likely to be {\it virtual} (cf. \cite{Bose2019}). 
They do not exist independently and, as such, they do
not contribute to the entropy, while the
interaction they cause is unitary. The physics near the black hole singularity is quite different. The huge mass concentration activates the genuine, ``non-virtual'', quantum degrees of freedom of the gravitational field.

In the presented analysis we have adopted a purely quantum information-theoretic approach, which is model-independent. Its advantage is to provide a general kinematic setup for all models of quantum gravity. One specific model of activation of the genuine quantum gravitational degrees of freedom was considered within loop quantum gravity \cite{Perez15,Perez17,Perez19}. A quite different mechanism involving wormholes was recently put forward in the context of string theory and the AdS/CFT correspondence \cite{Almheiri20,Maldacena2020,MarolfMaxfield2020}. 

In summary, we believe that the presented analysis offers a compelling explanation of the black hole information paradox and opens a new avenue towards quantum gravity.

\section{Acknowledgements}
 We thank L\'arus Thorlacius, Angelo Vulpiani and Rui Li for
  discussions and useful remarks. We thank Arun Kumar Pati and Sougato Bose for comments on the preprint of this work and for drawing our attention to refs. \cite{NoHiding} and \cite{Bose2019}, respectively. We also thank the anonymous Referees for their pertinent comments.
  
  This work was supported by the International Center for the Theory of Quantum Technologies
  (ICTQT, Gda\'nsk, Poland)
and by the Foundation for Polish Science through TEAM-NET project (contract no. POIR.04.04.00-00-17C1/18-00) (EA). 
The work of ME was supported by the National Science Centre in Poland under the research grant Sonatina (2017/24/C/ST2/00322).
The work of PH was supported  by the Foundation for Polish Science through IRAP project 
co-financed by the EU within the Smart Growth Operational Programme (contract no. 2018/MAB/5).

\end{document}